# A Utility-Based Channel Ranking for Cognitive Radio Systems

ADNAN QUADRI
Department of Electrical Engineering
University of North Dakota
Grand Forks, ND, USA, 58203
adnan.quadri@und.edu

*Abstract:* Growing number of wireless devices and networks has increased the demand for the scarce resource, radio spectrum. Next generation communication technologies, such as Cognitive Radio provides a promising solution to efficiently utilize radio spectrum whilst delivering improved data communication rate, service, and security. A cognitive radio system will be able to sense the availability of radio frequencies, analyze the condition of the sensed channels, and decide the best option for optimal communication. To select the best option out of the overwhelming amount of information, a channel ranking mechanism can be employed. While several channel ranking techniques have been proposed, most of them only consider the occupancy rate of the sensed channels. However, there are other significantly important parameters that provide information on the condition of channels and should also be considered during the ranking process. This paper proposes a utility-based channel ranking mechanism that takes into account signal-to-noise ratio and the occupancy rate of the channels to determine their usefulness or preference. The paper at first discusses the need for channel ranking and the involved process. Then the suitability of different mathematical functions is investigated for utility modeling of the channel based on its SNR and occupancy. Finally, results are provided that show improved channel ranking compared to that of spectrum occupancy based ranking.

## 1 Introduction

During the operation of a Cognitive Radio (CR) system, usually referred to as the cognitive cycle [1-2], one of the main steps is spectrum sensing, followed by spectrum analysis, and then spectrum decision making. A number of techniques to sense the radio spectrum has been proposed. These techniques can be classified into two categories: narrowband and wideband. Narrowband techniques aim to sense one frequency channel and include energy detection technique, cyclostationary features detection, matched filter detection.

Energy detection based spectrum sensing [3-8] measures the energy of the received signal samples. The computed energy level of the signal is then compared to a predetermined threshold. If the signal energy is above that threshold, the primary user signal is considered to be present, which implies that the sensed signal is occupied or unavailable for data communication by secondary users. Energy detection is one of the simplest and primitive techniques used for spectrum sensing. However, this technique is inefficient in noisy environments and is not able to distinguish between signals and noise [7].

At low SNR values, fixed threshold values usually fail to detect any primary user signal. To improve the energy detection technique, the authors of [4, 8] proposed ways to dynamically change the threshold and improve the detection of the PU signals. Unlike energy detection, cyclostationary feature [9-14] detection performs better in low SNR conditions as the technique detects primary user signals based on the correlation of the signal with its shifted version. Because noise is uncorrelated, this technique is able to distinguish between signals and noise. However, cyclostationary feature detection is more complex to implement and requires a high number of samples [9, 15]. Matched filter detection [16 -17] based signal detection requires prior knowledge of the PU signals. This technique uses pilot samples that are matched with samples of the received signal for the detection of the primary user. Although this technique does not require a large number of samples the need for prior knowledge of the PU signal is a major disadvantage.

Wideband spectrum sensing techniques aim at sensing a wide frequency range that includes one or several bands. To perform wideband spectrum sensing, the spectrum is divided into several sub-bands that are sensed either sequentially or simultaneously using one of the aforementioned sensing techniques. Examples of these techniques include 1-bit compressive sensing and multi-bit compressive sensing approaches [18-22].

To estimate the occupancy, two techniques are used: Frequentist and Bayesian inferences [23-27]. In the case of the Frequentist inference [23], the probability of an event to occur is inferred based on the frequency of occurrence of the event, provided that the event is observed for many trials. In the case of Bayesian inference, the probability of an event is inferred based on the previous observations and as well as the current ones. Bayesian inference is based on Bayesian Network which are probabilistic models that handle uncertainty. In [24], the authors proposed a simplified Bayesian inference model for spectrum occupancy that takes into account both deterministic and measured variables. In [27], the simplified model is further improved by including random variables, such as the probability of detection and false alarm of the sensing technique. Both the inference techniques aim to reduce uncertainties involved in channel occupancy measurement. However, Bayesian models allow the measurement of the occupancy in real time and take into consideration all or some of the variables that affect the occupancy, such as the characteristics of the sensing technique (detection, false alarm, and miss-detection probabilities), which increases the accuracy of estimation compared to Frequentist inference.

Other than spectrum occupancy, CR system must be able to analyze the condition of all the available channels using various other channel quality parameters. Some of the parameters include Signal-to-Noise Ratio (SNR), Signal-to-Interference Ratio (SINR), different types of delay associated with a channel, capacity of channels, and Bit Error Rate (BER) [28-33]. In [28][30], the authors discuss several BER estimation techniques and an estimation technique based on pilot samples are proposed, respectively. Similarly, in addition to BER, SINR can also be used and may provide better information on channel condition as the parameter considers the impact of interference during communication. For instance, in [32], a Bayesian approach is used to estimate and model SINR. The proposed technique reduces uncertainty in the estimation and provides better real-time measurements. In [33], the authors proposed a sample covariance matrix based SNR estimation, where an evolutionary algorithm is used to improve the accuracy of estimation.

Once the spectrum sensing and analysis are performed, a CR system goes through the decision-making phase, when the best channel for transmission is determined. Channel ranking mechanism assign ranks to the sensed channels, which enables the CR to efficiently utilize scarce radio spectrum while meeting certain communication requirements, such as quality of service, security, and latency. Several mechanisms have been proposed, where channels are ranked based on the primary user activity and state predictions [34-41]. In [36], a learning strategy for distributed channel selection in cognitive radio networks is proposed. The strategy considers different QoS requirements of CR systems/secondary users and means availability of channels in a network to determine the rank-optimal channels. Similarly, the authors of [34 - 35] determine the best channel for communication by estimating the occupancy rate. In [37 - 42], channel state prediction is performed by using predictive models and inference techniques, such as Bayesian inference. Based on the prediction of channel idle time and the accuracy of sensing, a secondary user then ranks the channel with the objective to use channels for a longer time.

Almost all the techniques discussed above use spectrum occupancy as a parameter to rank channels for data communication. Measuring spectrum occupancy alone is not enough and also does not indicate the quality of the sensed spectrum bands. However, SNR and spectrum occupancy rate together can be two QoS parameters that can be used to decide which channel is the most appropriate for data communication. Spectrum occupancy rate and SNR provide information on how readily available a channel is and how noisy is the radio frequency environment.

The process of selecting the best channel among the sensed channels requires assigning a score or ranking levels, which can be achieved by estimating the usefulness i.e. utility of the channel based on its SNR and occupancy. The next section outlines the process of modeling channel utility based on channel's SNR and occupancy rate. Some constraints and ideal scenarios that are considered to define the channel utility are also discussed in the following section.

## 2 Methodology: Utility-Based Channel Ranking

Utility modeling allows optimizing resource allocation, such as transmission power and modulation schemes of wireless communication systems by quantifying the usefulness of the resources [43]. Based on the usefulness of resources, ranking levels or scores can be assigned to indicate a preference for specific resources over the others. Therefore, utility-based modeling of communication parameters can help a CR system to decide the best course of action. A utility model based channel access has been proposed by [44] to enable cognitive radio systems to access a channel that can be used for a longer period of time and maintain a reasonable throughput before it has to be handed back to the primary or licensed user. Similarly, the authors of [45] propose an opportunistic channel selection in IEEE 801.11 based wireless mesh network by employing a utility modeling of the mesh network's load in different situations, which is then forwarded to a learning algorithm for the selection of the best channel. In [46-47], a utility-based resource allocation is applied to decide on the optimal transmission power allocation. Utility-based channel selection applied in the stated works depend on probabilistic models, which helps determine the future conditions of a channel by modeling collision probability, interference, and other metrics [48]. Some popular methods to design utility functions is the weighted-sum approach, linear-logarithmic or Cobb Douglas utility function, and constant-elasticity-of-substitution [49]. In the weighted-sum approach, the utility function of several objectives is added and their individual preference is dictated by the assigned weights. Similar to the weighted-sum approach, the linear-logarithmic utility assumes additivity but is found to be more useful as the logarithmic function is used to shape the utility.

In the case of channel ranking, the two parameters that are considered to determine the utility of a channel are SNR and spectrum occupancy, which are observed to show substitutive and complementary effects between each other. In cases where the utility function needs to reflect the substitutive effects, constant-elasticity-of-substitution (CES) utility function allows determining the degree of elasticity between two parameters and their relationship. In the next section, the use of CES utility function is described and utility modeling of SNR and spectrum occupancy is discussed. The utility model for the sensed channels can be defined by combining the utility values of the corresponding SNR and occupancy of these channels. Before we delve into defining a utility function that takes into account the utility values of both SNR and occupancy of a channel, it is necessary to outline the preferences for most desirable channel conditions.

The following are four scenarios, where a channel usefulness or 'utility' can be defined based on its SNR and occupancy.
1. A channel would be undesirable/less useful if it has high occupancy rate and also high SNR. In this case, even with a good SNR, the channel is less reliable as it may be found occupied most of the time.
2. A channel with low SNR (beyond acceptable SNR level) but with low occupancy rate is also undesirable. For such channels, although the occupancy rate is low it is still undesirable as the occupancy measurements at low SNR condition tend to be unreliable, increasing the probability of false alarm.
3. A channel with high SNR but also with low occupancy rate is most useful/desirable channel. Here, SNR and occupancy rate exhibits substitutive effects, where we want SNR to take over and have more impact on the utility calculation. As a result, such a channel will be defined with higher preference or utility.
4. A channel with low SNR (above acceptable SNR level) and intermediary occupancy rate (40 – 60 % occupancy rate) are also desirable. In such cases, it is convenient to have the occupancy take over and have the most impact on utility calculations for the channel.

To acknowledge the substitutive and complementary effects of SNR and occupancy a utility function needs to be defined to allow one parameter to be substituted by the other.
The constant elasticity of substitution (CES) utility function [49], can be defined as:

$$U_{SNR,Occ} = w_{snr}^{(1-\sigma)} U_{SNR}^{\sigma} + w_{occ}^{(1-\sigma)} U_{Occ}^{\sigma} \quad (1)$$

Where, $w_{snr}$ and $w_{occ}$ are the weight factors for SNR and occupancy, respectively. $U_{SNR}$ and $U_{Occ}$ are utility values for SNR and spectrum occupancy, and $\sigma$ determines the constant elasticity of substitution, $\rho = \frac{1}{1-\sigma}$. The constant $\sigma$, which is the elasticity between the parameters SNR and occupancy, introduces the degree to which one parameter can substitute another. This elasticity can be changed and based on the analysis of our application, a proper value of elasticity can be defined from case to case [49]. More details about the appropriation of elasticity and weight factor are discussed in the later section, where simulation results are provided and discussed. For now, it can be stated that the CES utility function enables us to substitute SNR for Occupancy and vice-versa as required, which finally allows us to define channel utility value based on its corresponding SNR and occupancy rate. Some utility models are used to make hard decisions while others for soft decision-making purposes. Hard decision making refers to the binary representation of 1 for 'ON switch' and 0 for 'OFF switch', and soft decision making allows a transient period between 0 and 1. As shown in Fig. 1, when the signal-to-interference ratio (SINR) is characterized by a utility modeling as in [50], the utility should be 0 to represent SINR beyond the acceptable level marked by the defined threshold. Any SINR value above the threshold will be perceived as a utility of 1 or highest utility as that kind of interference has no significant impact on the quality of service requirement of a communication system.

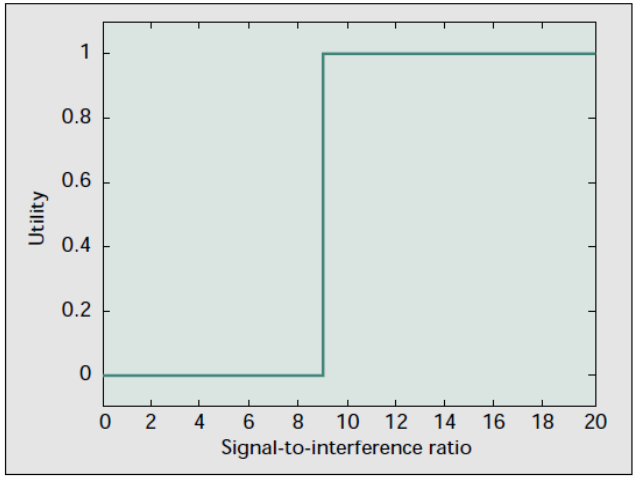

Fig. 1. Utility modeling of SINR to determine Quality of Service [50]

In this work, several mathematical functions are applied to define a utility function that appropriately characterizes our preference for higher SNR and lower occupancy. At first, varying SNR starting from negative 30 dB to 30 dB is considered. Different functions are defined and the corresponding utility is estimated for the considered range of SNR. In this case, the functions are defined to return utility values in the range of 0 to 100, where a utility value of 100 represents a preference for highest SNR conditions. To allow soft decision-making capabilities, sigmoid curve or logistic function appears to be useful as it renders utility values which represent high SNR, intermediate SNR values (SNR between 5 to negative 5 dB), and poor SNR conditions below negative 10 dB. The slope of the logistic function provides enough transient state to be able to make a soft decision by having a wide range of utility values for the considered range of SNR. Below are few utility functions that are used to represent SNR:

$$U_{SNR} = A/(1 + e^{-\alpha(X - X_o)}) \quad (2)$$
$$U_{SNR} = A(\frac{e^{\alpha X}}{1 + e^{\alpha X}}) \quad (3)$$
$$U_{SNR} = X_{max}(1 + tanh(\alpha(X))) \quad (4)$$
$$U_{SNR} = \frac{1}{2} + \frac{1}{2}(tanh(X/2)) \quad (5)$$

Where, $X_{max}$ represents the highest SNR value in the range of SNR values considered in the simulation, $X_o$ is the SNR value considered to be the midpoint for the sigmoid curve, and $\alpha$ determines the steepness of the curve and $A$ maximum value for utility. These four functions are continuous and render a utility between 0 and $A$, as SNR values ranges from $X_{min}$ to $X_{max}$. The utility model from these functions all appear to be 'S' shaped as logistic functions should be, which allows us to define wide ranges of utility values representing SNR in dB. Utilizing the symmetry property of logistic functions, same but reversed equations can be used to represent the utility of spectrum occupancy, where $Y$ is occupancy rate, $Y_{max}$ is a constant that is the highest occupancy rate, and $Y_o$ is the occupancy rate considered to be the midpoint for the sigmoid curve. As lower occupancy is preferred, reversed sigmoid curve provides higher utility for low occupancy and lower utility for high spectrum occupancy rates.

## 3 Results & Discussion

Fig. 2 illustrates the previously defined utility functions for a fixed range of SNR values from -20dB to +20 dB. MATLAB is used as the platform to implement the simulations. For the experiments, $\alpha$ that determines the steepness of the curve is defined to be 0.1 for the hyperbolic tan function and 0.2 for the logistic functions. Maximum value for utility $A$ is defined to be 100, so that the utility values are in the range of 0 to 100. $X_{max}$, which is the highest SNR value in the fixed range of SNR values, comes up to be 20 dB for this experiment. Subsequently, $X_o$ that is midpoint of the range of SNR is 0 dB, in this case. As seen from Fig. 2, the 'S' shape of (2), the logistic function, allows to define the higher utility values for high SNR conditions and lower utility values for low SNR conditions. However, Equation 5 renders a utility model that is appropriate for the case of hard decision making, as the utility values see a sharp rise and fall for any SNR values above and below -5 and 5 dB. Excluding Equation 5 from consideration, rest

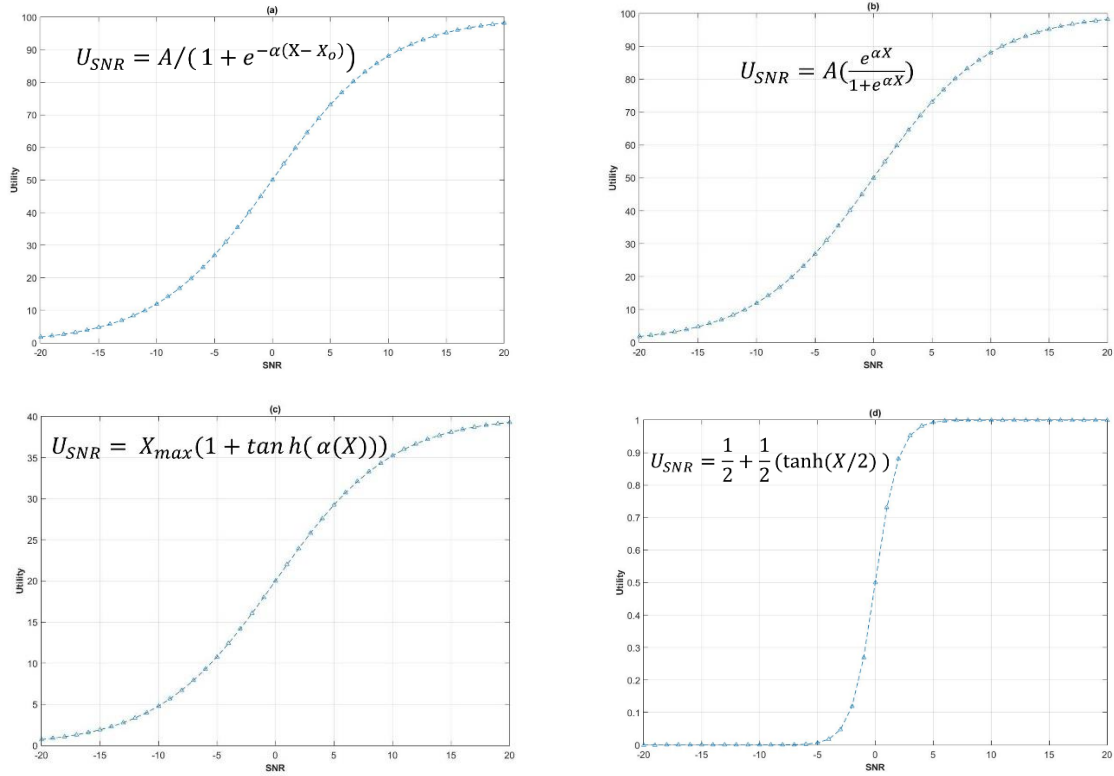

Fig. 2. (a) Variant of the logistic function, (b) Logistic function, (c) Hyperbolic tangent function scaled by maximum SNR value, (d) Logistic function as a scaled hyperbolic tangent function

of the simulation illustrating utility values over occupancy will narrow down the choice of the most accurate utility function to model both occupancy and SNR for a given channel.

The same utility functions are then used to model the utility for the spectrum occupancy rate, which ranges from 0 to 1. Fig. 3 illustrates the utility values over occupancy for each of the four utility functions. For this simulation, α that determines the steepness of the curve is defined to be 5 for all the utility functions except for Equation 5, which is defined to have 0.5. The maximum value for utility A is defined to be 100. As seen in the figure, the 'S' shape is not retained anymore by both Equations 2 and 5, the logistic and hyperbolic tan functions. This is due to the range of occupancy values being under 1. However, the utility modeling still characterizes the lower occupancy rates with higher utility values, which is desired in our case. When compared from the figure, functions (2), (3) and (4) renders similar utility model, where (2) and (3) have different maximum utility but similar steepness, (4) has a steeper descent and a maximum utility at 1. Figure 4(d), which is generated from Equation (5) renders utility values that follow a linear relationship between utility and corresponding occupancy.

Based on the results of the previous experiments, Function (5) offers the most suitable utility model as it characterizes better SNR and occupancy with high utility values and reprimands degrading conditions with lower utility. This utility model, defined by the utility function (5), allows soft-decision making capabilities, where the transition from good to a worse condition doesn't follow steep descent. Once the utility for SNR and occupancy is estimated, constant elasticity of substitution defined in (1) is used to estimate the combined utility of the corresponding channel.

Table I and II provides utility based channel ranking and occupancy based channel preference. Table I shows the combined utility of all the sensed channels based on their corresponding utility value of SNR and occupancy. Channels are then ranked in a descending order based on their utility values. As seen from Table I, the highest ranked channel is the one with the highest utility value and has a reasonable SNR and occupancy rate compared to channels with lower utility. When compared to to the first channel in Table II, it is observed that the channel with the lowest occupancy is selected although the corresponding SNR is lower than that of the second channel in Table II. The utility based channel selection is able to recognize the two channels with same occupancy rate but different SNR conditions, where ranked 1 channel in Table I has slightly better SNR.

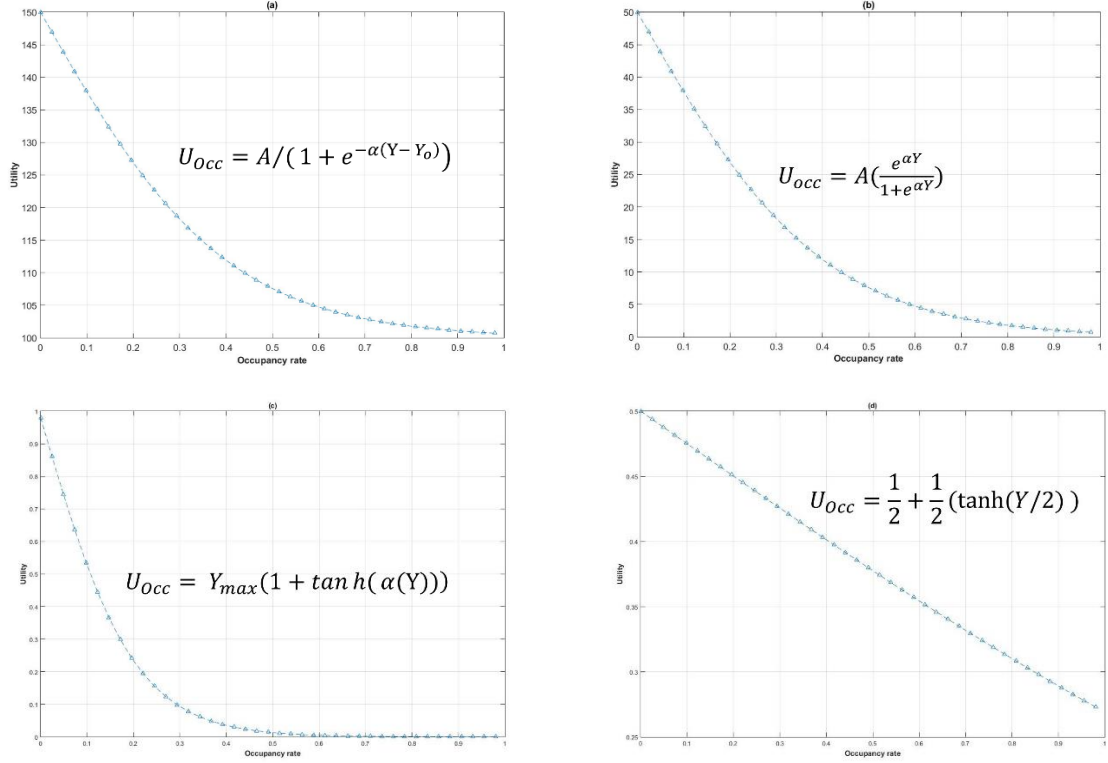

Fig. 3. (a) Variant of the logistic function, (b) Logistic function, (c) Hyperbolic tangent function scaled by maximum SNR value, (d) Logistic function as a scaled hyperbolic tangent function

Similarly, the yellow colored row in Table II shows the channel ranked 18 to be preferred when only the channel occupancy is considered. Red colored rows in Table II are the channels that should have been ranked low as they have bad SNR conditions. In Table I, the red colored channels are ranked 27 and 24 due to their degrading SNR values. From these two tables, it can be deduced that utility-based channel ranking and selection helps to recognize and perform a tradeoff between SNR and spectrum occupancy rate to prefer better channels than that of the occupancy based channel selection method. Therefore, it can be concluded that the utility-based method provides better means of decision-making for a CR system to select the best channels among all the sensed channels.

The utility-based channel ranking technique, proposed in this chapter is computationally simple compared to other channel ranking mechanisms, which involve the implementation of complex algorithms requiring a large number of iterations and multiple steps [51-55]. The CES function relies on two important communication parameters, SNR, and spectrum occupancy rate, which provides important information regarding the channel condition. The CES based utility function can also be used with other parameters or channel quality metrics, such as Bit Error Rate (BER) and Signal-to-Interference Ratio (SINR) along with spectrum occupancy.

Table I: Utility-based channel ranking

| Rank | Frequency (GHz) | SNR | Occupancy (%) | Combined Utility - $U(\gamma, Occ)$ |
|---|---|---|---|---|
| 1 | 2.462 | 12 | 1 | 94 |
| 2 | 2.437 | 19 | 6 | 93 |
| 3 | 2.437 | 8 | 1 | 89 |
| 4 | 5.765 | 11 | 9 | 84 |
| 5 | 5.765 | 17 | 12 | 84 |
| 6 | 2.462 | 8 | 6 | 83 |
| 7 | 1.88 | 17 | 14 | 82 |
| 8 | 2.462 | 18 | 15 | 81 |
| 9 | 1.88 | 17 | 15 | 81 |
| 10 | 2.412 | 13 | 16 | 77 |
| 11 | 5.765 | 12 | 16 | 76 |
| 12 | 1.88 | 7 | 13 | 73 |
| 13 | 2.462 | 19 | 24 | 71 |
| 14 | 2.412 | 8 | 28 | 59 |
| 15 | 2.437 | 5 | 24 | 58 |
| 16 | 1.88 | -1 | 11 | 56 |
| 17 | 2.437 | -2 | 10 | 54 |
| 18 | 1.88 | -2 | 10 | 54 |

Table II: Occupancy based channel selection

| Frequency (GHz) | SNR | Occupancy (%) | Ranked by Utility based ranking as |
|---|---|---|---|
| 2.437 | 8 | 1 | Ranked 3 |
| 2.462 | 12 | 1 | Ranked 1 |
| 5.765 | -17 | 1 | Ranked 27 |
| 5.765 | -12 | 5 | Ranked 24 |
| 2.437 | 19 | 6 | Ranked 2 |
| 2.462 | 8 | 6 | Ranked 6 |
| 5.765 | 11 | 9 | Ranked 4 |
| 2.412 | -18 | 10 | Ranked 34 |
| 2.437 | -2 | 10 | Ranked 17 |
| 2.462 | -6 | 10 | Ranked 22 |
| 1.88 | -2 | 10 | Ranked 18 |
| 1.88 | -1 | 11 | Ranked 16 |
| 5.765 | 17 | 12 | Ranked 5 |
| 1.88 | -9 | 12 | Ranked 25 |
| 1.88 | 7 | 13 | Ranked 12 |
| 1.88 | 17 | 14 | Ranked 7 |
| 2.437 | -17 | 15 | Ranked 37 |
| 2.462 | 18 | 15 | Ranked 8 |
| 1.88 | 17 | 15 | Ranked 9 |

## 4 Conclusions

In this chapter, functions for channel ranking mechanism for cognitive radio systems are discussed. Simultaneously, several communication parameters that provide information on channel conditions are also studied and their impact on channel selection mechanism is discussed. The utility modeling for two important channel condition parameters, SNR and spectrum occupancy is then provided. CES utility function is defined to model channel utility, which combines the utility values of channel's SNR and occupancy rate. Subsequently, the channel utility model is used to rank the sensed channels.

Simulations were performed for multiple frequencies, ranging from megahertz to gigahertz, with different corresponding SNR and occupancy rates. Results indicate that the proposed utility-based channel ranking performed better with increased accuracy in ranking optimal channels for communication, compared to the usual occupancy-based channel selection by CR systems.